# Telecommunication band InAs quantum dots and dashes embedded in different barrier materials

Nahid A Jahan,[1,2] Claus Hermannstädter,[1] Jae-Hoon Huh,[1] Hirotaka Sasakura,[1] Thomas J Rotter,[3] Pankaj Ahirwar,[3] Ganesh Balakrishnan,[3] Kouichi Akahane,[4] Masahide Sasaki,[4] Hidekazu Kumano[1] and Ikuo Suemune[1]

[1] Research Institute for Electronic Science, Hokkaido University, Sapporo 001-0021, Japan
[2] Graduate School of Information Science and Technology, Hokkadido University, Sapporo 060-0814, Japan
[3] Center for High Technology Materials, University of New Mexico, Albuquerque, New Mexico 87131, (505) 277-0111, USA
[4] National Institute of Information and Communications Technology, Koganei, Tokyo 184-8795, Japan

E-mail: nahid@es.hokudai.ac.jp, claus@es.hokudai.ac.jp

**Abstract.** We investigate the long wavelength (1.2 to 1.55 μm) photoluminescence of high-density InAs quantum dots and dashes, which were grown on InP substrates. We analyze the temperature dependence of the recombination and carrier distribution on the alloy composition of the barrier materials, $In_xGa_yAl_{1-x-y}As$, and on the existence of a wetting layer. Carrier escape and transfer are discussed based on temperature dependent photoluminescence measurements and theoretical considerations about the heterostructures' confinement energies and band structure. We propose two different contributions to the thermal quenching, which can explain the observations for both the quantum dot and dash samples. Among these one is a unique phenomenon for high density quantum dot/dash ensembles which is related to significant inter-dot/dash coupling. With the goal ahead to use these dots and dashes for quantum optical applications on the single-dot/dash level in the telecommunication C band as well as at elevated temperatures we present first steps towards the realization of such devices.

# 1. Introduction

Numerous applications of semiconductor quantum dots (QDs) have been reported throughout the past decades. They have been widely used as promising and favourable active media in semiconductor QD lasers [1—4], both as ensembles of QDs of the highest possible densities and as single QDs (SQDs) [5—6], and also as photo detectors [7]. Microscopic spectroscopy studies of individual semiconductor QDs have shown that SQDs have excellent potential as the heart of single photon sources [8—10] and sources of entangled photon pairs [11—13] for high-speed quantum information and communication. With the focus on practical QD based quantum communication through fibre-optical networks, the attention is drawn to the telecommunication bands, especially focusing on the O band (0.912—0.984 eV; 1.360—1.260 μm) and the C band (0.792—0.810 eV; 1.565—1.530 μm). It has to be noted that on the SQD side the main part of the previous work was reported on spectral ranges which do not cover the telecommunication bands. This includes mainly the larger band gap range, e.g., around 1—1.5 eV (1200—800 nm) for the up to 7% lattice mismatched In(Ga)As/GaAs (III-V) QDs and around 1.8—3.5 eV (700—350 nm) for CdSe/ZnS (II-VI) QDs, all grown on GaAs substrates, as well as, the group-III-nitrides with their respective energies from 1 to 6 eV (1200—200 nm). This consequently hauled the interest towards InAs/InP heterostructures of smaller lattice mismatch (~3%), which cover the near-infrared wavelength range (0.953—0.799 eV; 1.3—1.55 μm) and thereby meet the suitability for silica based fibre networks.

Hereafter we want to introduce the following abbreviations that will be used throughout this manuscript: "QDots" for almost cylindrically symmetric quantum dots grown on InP(311)B substrates, "QDashes" for elongated dots grown on InP(001) substrates and the conventionally used "QDs" for both QDots and/or QDashes, without distinguishing between them, in a context where the exact definition is not particularly important.

The morphological properties of QDs crucially depend on the growth condition, substrate orientation and the nature of surrounding matrices. For the particular InAs/InP system, we can observe the formation of QDashes, especially when grown on InP(001) substrates [14—15]. While, InAs/InP QDs can have circular symmetry whereby grown on (311)B InP substrates [16—17]. Moreover, the growth of telecom band InAs QDs on InP tends to result in high density QDs (~$10^{11}$cm$^{-2}$). This can provide the possibility of inter-dot tunnel coupling among neighbouring QDs. Although, a number of studies on carrier transfer [18—19] have been performed on InAs QDs grown on GaAs, very few experimental studies have been carried out on carrier tunnelling mechanism in such high-density InAs QDs grown on InP substrates. The optical gaps or confinement energies of QDs, i.e., the energy differences between confined states and continuum states, such as a wetting layer (WL) or barrier, are important quantities, too. This is especially the case when the system's temperature is increased above the liquid Helium temperature of 4.2 K. In order to achieve higher confinement energies, the inclusion of a variety of higher-bandgap barrier materials was suggested and reported [20—23]. For group-III-As/P hetero-structures, which are treated in this work, the highest energy gaps can be realized by increasing the Al content of the barrier material, which is, among others, discussed in detail by Schulz et al. in ref. [23]. Another closely related topic is the growth mode of the QDs, with the possible consequence of the QDs coupling to a WL which can be energetically very close, such as for the typical Stranski-Krastanov growth [24] of self-assembled QDs, e.g., InAs on GaAs, where the related energy gap can be as small as some 10 meV [25—26]. Concerning the performances of photonic devices, the detailed understanding of the thermal quenching mechanism of QD luminescence is inevitable. Extensive study of thermal escape modelling has been reported in several publications, e.g. [27—29]. Despite this, there is no general agreement about the fundamental mechanisms behind the carrier thermal escape from confined QD states. Moreover, the carrier thermal escape can be sensibly influenced by the presence of a WL, whose existence can therefore be regarded as a crucial problem limiting the performances of QDs as a laser medium.

Based on the aforementioned topics, we discuss some of these key fundamental parameters of QD heterostructures in this paper, such as, QDs with different optical gap and confinement energies, which could be achieved by changing the QD sizes and barrier composition. In line with this, we study the related carrier escape and transfer mechanisms, the existence and impact of a WL, and, as a consequence of these, the temperature range in which different types of QDs can be used. We investigate the photoluminescence (PL) of the respective QDs as ensembles and as single emitters with the focus on the temperature dependent modification of the spectral function and total intensity. Based on our measurements, we conclude that no WL exists below our QDots grown on InP(311)B substrates. The activation energies related to the thermal quenching of the PL are determined and the mechanisms which might be responsible for this phenomenon are discussed and proposed. We show that our results are well explained by the quasi-correlated electron-hole escape mechanism. Finally, we also present a suitable way to use high density QDs as a source of single photons and present distinct single QDash emission at 1.55 μm.

## 2. Sample preparation of the QD heterostructures

### 2.1 QDots in $In_{0.53}Al_{0.22}Ga_{0.25}As$ barriers

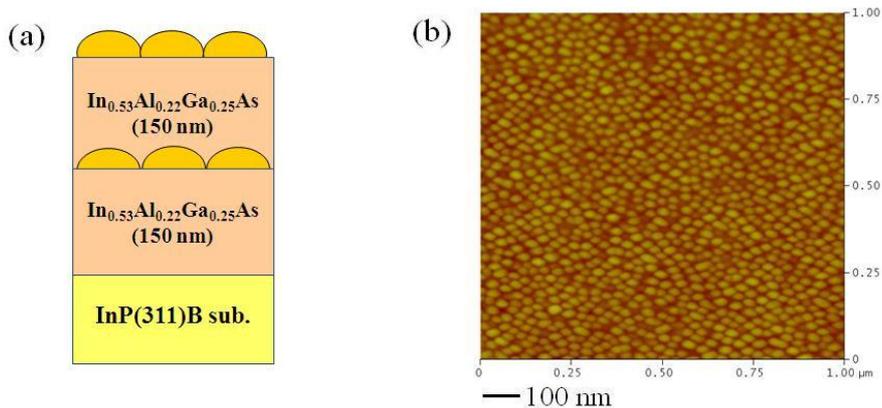

**Figure 1: (a)** Schematic illustration of the InAs/$In_{0.53}Al_{0.22}Ga_{0.25}As$ QDot heterostructure and **(b)** 1x1 μm$^2$ AFM image of the open dots on top of the 4 ML QDot sample.

The high density InAs QDots that we used for this study were grown by solid source molecular beam epitaxy (MBE) on a lattice-matched $In_{0.53}Al_{0.22}Ga_{0.25}As$ buffer on InP(311)B substrates, where a high dot density and homogeneity as well as a good rotational dot-symmetry can be achieved [30—33]. The samples feature an optically active layer of QDots with a nominal thickness of four to six monolayers (MLs) of InAs which is embedded between 150 nm thick $In_{0.53}Al_{0.22}Ga_{0.25}As$ barrier layers (figure 1 (a)). An identical QDot layer on top of the upper barrier layer was added for atomic force microscope (AFM) imaging of the samples. The AFM image conveys that the shape of the InAs QDots is almost rotationally symmetric with a very uniform spatial distribution (figure 1(b)); the areal dot density is around $1.1 \times 10^{11}$ cm$^{-2}$, the average lateral diameters of the QDots is 29 : 23 nm and the height of the uncapped open QDots is 3.4 nm.

*2.2 QDashes in $In_{0.52}Al_{0.48}As$ barriers*

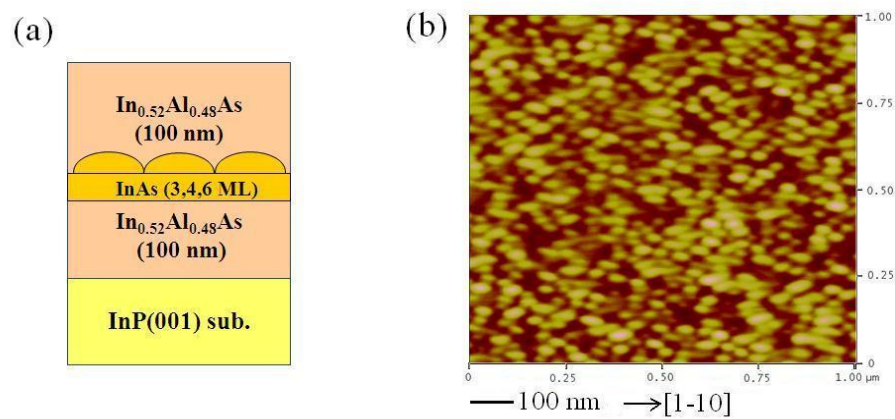

**Figure 2: (a)** Schematic illustration of the InAs/$In_{0.52}Al_{0.48}As$ QDash heterostructure, **(b)** 1x1 μm$^2$ AFM image of the 6 ML QDash sample, in this case a sample without the $In_{0.52}Al_{0.48}As$ cap layer was used for AFM imaging.

According to the study reported by Balakrishnan et al. in ref. [34], the QDash asymmetry, which depends on the relative surface migration along the [1-10] and the [110] is influenced by both strain and surface bonds. Zincblende or tetrahedral structures like InAlAs create a bonding asymmetry on the (100) surface in two directions. Group-V stabilized surfaces show the tendency to form islands elongated along the [1-10] direction while group-III stabilization leads to elongation along the [110] direction [34—35]. In the low strain growth mode, the directional property of the substrate is transferred to the island ensemble grown on it. There is a tendency of forming dash-like shapes observed in InAs dots grown on InAlAs/InP(001) [36]. In this work, the high density InAs QDashes samples were grown by MBE on a lattice matched $In_{0.52}Al_{0.48}As$ buffer on InP(001) substrates. Besides lattice matching, another incentive of introducing the high Aluminium concentration in the barrier layers is the increased confinement energy of the QDashes and thus the possibility to ensure that the charge carriers remain confined within the QDashes up to higher temperatures. The QDashes were grown by varying the nominal thickness of the optically active layer between three and six MLs of InAs (figure 2 (a)). For these particular InAs/$In_{0.52}Al_{0.48}As$ QDash samples, we observed a less homogeneous distribution of InAs islands as compared to the aforementioned QDots. Both dash and elongated dot-like shapes can be observed in the AFM image (figure 2 (b)). The observed 6 ML QDashes appear favourably elongated along the [1-10] crystal direction with an aspect ratio of approximately 2.5 : 1 (60 : 24 nm average lateral diameters), an average height of the uncapped QDashes of 1.88 nm and an overall areal QDash density of around $5.1 \times 10^{10}$ cm$^{-2}$.

## 3. Optical properties and discussion

*3.1 Measurement setup*

The low temperature continuous wave (cw) macroscopic PL measurements were performed under non-resonant excitation at low average power densities of 5—27 W/cm$^2$ using a frequency doubled Nd:YAG laser at 2.33 eV (532 nm). The temperature dependent PL spectra of the 4 ML QDot sample and all QDash samples were measured using a 75-cm monochromator with a 300 gr./mm grating and a liquid Nitrogen cooled InGaAs photodiode array detector. For the 6 ML QDot sample a slightly modified detector set-up was used because the PL energy exceeded the cut-off wavelength of 1.6 μm of the above mentioned detector. A 30-cm monochromator with a rotating 200 gr./mm grating and a liquid Nitrogen cooled InGaAs photodiode detector with an extended sensitivity range up to 2.2 μm was used in lock-in operation with the excitation laser being modulated by a mechanical chopper. The integration time was kept constant for all measurements, only the detector sensitivity was enhanced for the lowest signals levels at elevated temperature.

*3.2 Low temperature PL analysis: identification of active quantum states and their optical gaps*

The low temperature ensemble PL of 4 ML InAs QDots embedded in In$_{0.53}$Al$_{0.22}$Ga$_{0.25}$As barriers and InAs QDashes in In$_{0.52}$Al$_{0.48}$As barriers is displayed in figure 3.

Firstly, in figure 3 (a) the optical signatures of the barrier and substrate as well as the respective energy separations to the QDot PL are highlighted. The 4 ML QDot PL emerges at around 0.885 eV (1.401 μm), the emission centered around 1.142 eV (1.085 μm) is attributed to the In$_{0.53}$Al$_{0.22}$Ga$_{0.25}$As barrier, and the substrate emission raises at 1.425 eV (0.870 μm). The energy separation between the QDot ensemble and the barrier (substrate) is 257 meV (540 meV). No WL emission is observed from these QDot samples.

In a similar way, a detailed view of a typical spectrum from a QDash sample is shown in figure 3 (b), exemplified by the 4 ML QDashes. For the identification of active quantum states from the QDash sample, the excitation power density was increased by reducing the laser spot diameter. This modification was necessary because the QDash samples seem to exhibit faster barrier recombination less efficient carrier capture to the QDashes under non-resonant excitation. The result is a deviation from the ideal Gaussian shape in the QDash emission because of a reduced number of QDashes contributing to the PL when compard to the QDot spectrum in figure 3 (a). In contrast to the QDot samples, we observed the existence of a WL emissiom along with barrier and substrate emission in the QDash samples. The QDash PL emerges at around 1.005 eV (1.233 μm) and the WL PL at an energy of 1.377 eV (0.900 μm). Substrate emission is observed at 1.425 eV (0.870 μm) and barrier emission at 1.530 eV (0.810 μm). The PL spectrum again provides clear information about the energy difference between the 4 ML QDash confined states and other optically active higher energy states such as barrier, substrate and WL. The optical energy separation between QDash ensemble and WL is 372 meV. The energy difference between QDash and substrate emission is 420 meV and the energy separation between QDash and barrier emission is 525 meV.

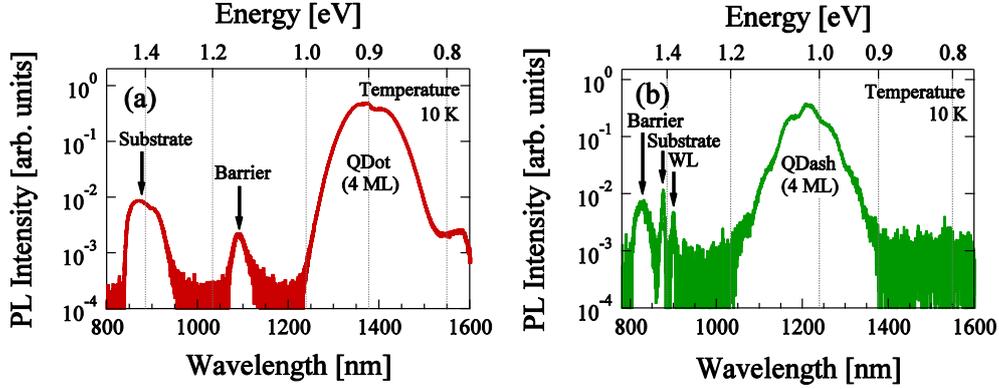

**Figure 3: (a)** Low temperature PL spectrum of 4 ML InAs/In$_{0.53}$Al$_{0.22}$Ga$_{0.25}$As QDots under low power non-resonant excitation featuring the QDot PL at 0.885 eV (1.401 μm), the barrier emission at 1.142 eV (1.085 μm) and the substrate emission at 1.425 eV (0.870 μm). **(b)** Low temperature PL spectrum of 4 ML InAs/In$_{0.52}$Al$_{0.48}$As QDashes under medium power non-resonant excitation featuring the QDash PL at 1.005 eV (1.233 μm) as well as emission from WL (1.377eV; 0.900 μm) substrate (1.425eV; 0.870 μm) and (1.53 eV; 0.810 μm).

*3.3 Temperature dependent PL analysis*

The temperature dependent PL spectra of the 4 ML QDots shown in figure 4 (a) was obtained under low power cw excitation. The dip in the vicinity of 0.885 eV (1.401 μm) is due to the OH$^-$ absorption in the fibres used for these measurements. Between 10 and 200 K, we can observe that the QDot ensemble PL successively red-shifts and eventually decreases; a more detailed analysis follows in the next section and in figure 8. The temperature dependent PL spectra of the 6 ML QDots are shown in figure 4 (b), where the low temperature spectrum exhibits the QDot PL centred at around 0.800 eV (1.550 μm). The 6 ML QDot spectra are shown for temperatures between 10 and 280 K.

To study the thermal quenching of carriers at elevated temperature, understanding of the temperature dependent PL is an inevitable issue. Since the general description and proper understanding of thermally activated carrier escape from QD ground states to some energetically higher states is the key issue, the carrier (re)distribution and escape mechanisms are discussed quantitatively in this section. We can deduce from the temperature varied PL of QDots, shown in figure 4, that the integrated intensity is almost constant up to 40 K and exhibits a small reduction between 40 and 70 K. Above 70 K the intensity starts to diminish speedily, indicating the presence of non-radiative recombination mechanisms.

The integrated PL intensities of 4 ML and 6 ML QDots are plotted as a function of inverse temperature in figure 4 (c) and (d). Both samples clearly show the two aforementioned temperature regimes, which are treated as two thermally activated non-radiative recombination mechanisms.

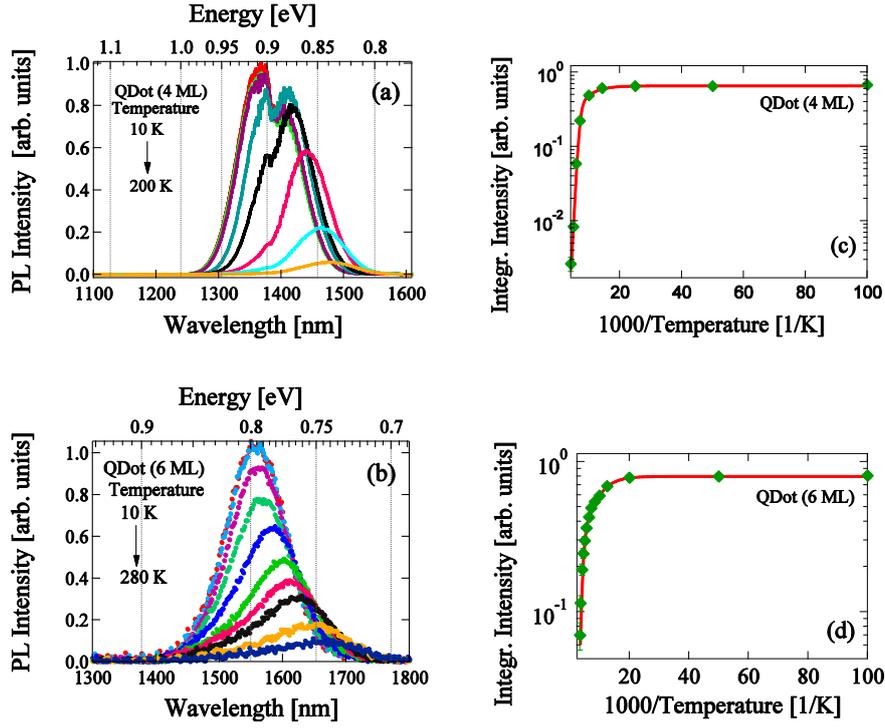

**Figure 4:** (a) Temperature dependent PL spectra of 4 ML QDots at T = 10, 20, 40, 70, 100, 140, 170, and 200 K under non-resonant excitation (6.7 W/cm$^2$). (b) PL spectra of 6 ML QDots at T = 10, 20, 50, 80, 120, 160, 180, 200, 240, and 280 K under non-resonant excitation (27 W/cm$^2$). The integrated PL intensity for the (c) 4 ML and (d) 6 ML InAs QDots are displayed as a function of inverse temperature. The fit curves (red solid lines) were obtained using equation (1).

The solid lines result from the best fit to the experimental data by the Arrhenius-type equation [37]

$$I(T) = \frac{I_0}{1 + B_1 \exp\left(-\frac{E_{a1}}{kT}\right) + B_2 \exp\left(-\frac{E_{a2}}{kT}\right)} \quad . \quad (1)$$

$I_0$ is the integrated PL intensity at 0 K, $E_{a1}$ and $E_{a2}$ are the thermal activation energies, $B_1$ and $B_2$ are dimensionless coefficients, $k$ is the Boltzmann constant, and $T$ is the sample temperature.

For the 4 ML InAs QDots [figure 4 (c)], the best fit using equation (1) yields the activation energies $E_{a1}$ = 22 meV and $E_{a2}$ = 157 meV and the corresponding coefficients $B_1$ = 4.7 and $B_2$ = 6.4 x 10$^5$. Similarly, for the 6 ML QDots [figure 4 (d)], we obtained activation energies of $E_{a1}$ = 20 meV and $E_{a2}$ = 189 meV and the corresponding coefficients $B_1$ = 4.3 and $B_2$ = 2.4 x 10$^4$.

The PL of InAs QDashes was investigated as a function of temperature in a comparable way to the QDots and is displayed in figure 5. The temperature dependence of the 3 ML QDash PL spectra between 10 and 220 K is displayed in figure 5 (a); similarly, in figures 5 (b) (5 (c)) the temperature dependent PL spectra of 4 ML (6 ML) QDashes between 10 and 293 K (250 K) are shown.

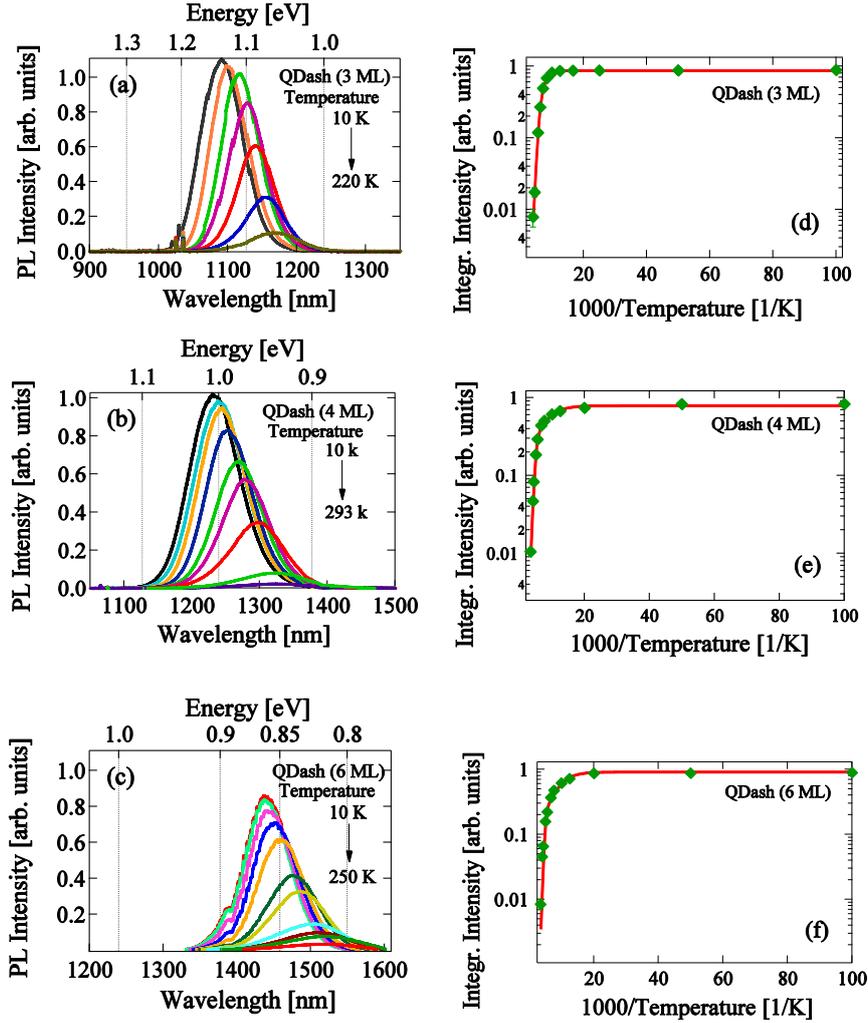

**Figure 5:** **(a)** Temperature dependent PL spectra of 3 ML QDashes at T = 10, 60, 100, 140, 160, 180, and 220 K using non-resonant excitation (13 W/cm$^2$). The low temperature 3 ML QDashes PL shows emission wavelength arising at 1.13 eV (1.095 μm). **(b)** The temperature dependent macroscopic PL spectra of 4 ML QDashes were recorded under non-resonant excitation (13 W/cm$^2$) at T = 10, 50, 80, 100, 120, 140, 180, 240, and 293 K. The low temperature center wavelength of the 4 ML QDashes PL is observed around 1.005 eV (1.233 μm). **(c)** The PL measurement of 6 ML QDash as a function of temperature at T = 10, 20, 50, 80, 100, 130, 150, 180, 200, 230, and 250 K under non-resonant excitation (13 W/cm$^2$) with the PL peak wavelength around 0.861eV (1.44 μm). The integrated PL intensity for the **(d)** 3 ML **(e)** 4 ML and **(f)** 6 ML InAs QDashes are displayed as a function of the inverse temperature. The fit curves (red solid lines) were obtained using equation (1).

The low temperature (10 K) PL spectra of the 3, 4 and 6 ML QDashes are centered around 1.13 eV (1.095 μm), 1.005 eV (1.233 μm) and 0.861 eV (1.440 μm), respectively. This implies that the QDash emission exhibits an increasing energy separation from the WL, barrier and substrate with increasing QDash size (number of MLs of the QDash layer).

The integrated PL intensities were plotted as a function of inverse temperature for all three QDash samples (3—6 MLs) and are shown respectively in figure 5 (d), (e) and (f). The experimental data were fitted in the same way as discussed above using equation (1). The best fits again include the contribution of two non-radiative recombination mechanisms. For the 3 ML QDashes the best fit results in two activation energies $E_{a1}$ = 25 meV and $E_{a2}$ = 180 meV and the corresponding coefficients $B_1$ = 4.4 and $B_2$ = 4.8x $10^5$ (figure 5 (d)). For the 4 ML QDashes the activation energies $E_{a1}$ = 17 meV and $E_{a2}$ = 224 meV and coefficients $B_1$ = 2.8 and $B_2$ = 2.4 x $10^6$ are obtained (figure 5 (e)), and for the 6 ML QDashes $E_{a1}$ = 19 meV, $E_{a2}$ = 275 meV, $B_1$ = 3.4 and $B_2$ = 1.7 x $10^7$ (figure 5 (f)).

The activation energies, fitting coefficients and the one-standard deviation fit errors of all the samples are listed in table 1. The activation energies of QDots and QDashes are also displayed in figure 6 with respect to their corresponding number of MLs. The closed diamonds represent activation energies of QDashes, while the closed circles represent QDot activation energies. The solid and dashed lines depicted in figure 6 will be discussed later conjointly in section 3.4.3 and section 3.4.4.

**Table 1.** The list of activation energies and fitting coefficients

| Sample | Activation Energy | | Fitting Coefficient | |
|---|---|---|---|---|
| | $E_{a1}$ (meV) | $E_{a2}$ (meV) | $B_1$ | $B_2$ |
| 4 ML QDots | 22 ± 2.2 | 157 ± 4.8 | 4.7 ± 0.41 | 6.4 x $10^5$ ± 0.09 x $10^5$ |
| 6 ML QDots | 20 ± 1.4 | 189 ± 6.0 | 4.3 ± 0.71 | 2.4 x $10^4$ ± 0.06 x $10^4$ |
| 3 ML QDashes | 25 ± 3.0 | 180 ± 7.3 | 4.4 ± 0.51 | 4.8 x $10^5$ ± 0.09 x $10^5$ |
| 4 ML QDashes | 17 ± 2.0 | 224 ± 5.8 | 2.8 ± 0.36 | 2.4 x $10^6$ ± 0.05 x $10^6$ |
| 6 ML QDashes | 19 ± 2.8 | 275 ± 3.7 | 3.4 ± 0.57 | 1.7 x $10^7$ ± 0.04 x $10^7$ |

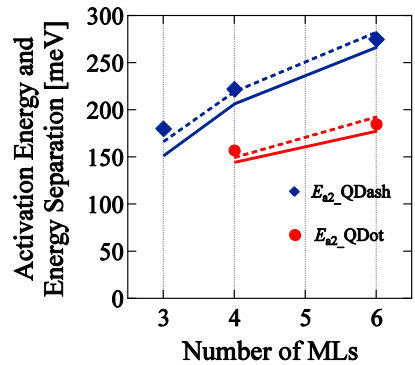

**Figure 6:** Activation energies of QDots and QDashes as a function of the number of MLs of the respective QDs. The blue diamonds ($E_{a2}$_QDash) represent the QDash activation energies. The red circles ($E_{a2}$_QDot) display QDot activation energies. The blue (red) dashed line represents one half of the energy difference between the low energy edge of QDash (QDot) PL and the WL (barrier) emission energy determined at low temperature for the corresponding number of MLs. The solid blue (red) line displays one half of the energy separation between QDash (QDot) PL peak (centre) energy and WL (barrier) energy deduced at high temperatures including the additional PL peak shift described in section [3.4.4].

*3.4 Discussion of thermal activation processes*

*3.4.1 Fitting coefficients $B_{1,2}$*. The fitting coefficients $B_{1,2}$ listed in table 1 are defined simply as the ratio of carrier capture time of the QDs and carrier escape time of the QDs by the respective non-radiative process, which is narrated in a distinctive manner by Le Ru et al. in ref. [38]. Consequently a scenario with a significantly faster carrier capture than escape time leads to small values of $B_{1,2}$ and vice versa. From the experimental results of both samples, it is obvious that the value of $B_1$ is significantly smaller when compared to $B_2$. Hence the term scaled by $B_2$ is the dominating contribution to the thermally activated carrier escape process, while the term scaled by $B_1$ plays a minor role.

*3.4.2 Activation energies $E_{a2}$*. First, we discuss the mechanism behind the activation energies $E_{a2}$. The following ratio,

$$v_i = E_{a2,i}/\Delta E_i \ , \qquad (2)$$

is introduced, where, $E_{a2,i}$ is the activation energy and $\Delta E_i$ is the energy difference of the transition energies between the QDots and any associated higher energy state (e.g. the barrier or WL) which is designated by the index $i$.

This ratio leads to the three possible scenario of carrier escape. Firstly, for $v_i = 1$, an escape process of fully correlated electron-hole pairs, i.e., for bound excitons is described. Secondly, for $v_i = \frac{1}{2}$, an escape process where $E_{a2,i}$ is exactly $\frac{1}{2}\Delta E_i$ is described. According to Yang et al. [39] and Michler et al. [40], this condition is derived for the escape process of electrons and holes with equal probability. This situation is most probable when the band discontinuities in the conduction and valence bands are similar and the electron and hole concentrations are equal.

Thirdly, for $v_i < \frac{1}{2}$, an escape mechanism is described which is dominated by single carrier escape with $E_{a2,i}$ corresponding to the less confined carriers due to the smaller discontinuity of either conduction or valance band.

We have displayed the values of $\frac{1}{2}\Delta E_i$, i.e., one half of the values of the energy differences between the QDashes (QDots) emission energies and the WL (barrier) emission energy for the respective number of MLs, by the blue (red) dashed lines in figure 6. The energy differences were determined using the low-temperature PL spectra of all the samples. The low energy edge of the QDash (QDot) PL, defined by $I_{max}/(2e^2)$, was used to determine the energy differences, where $I_{max}$ is the maximum intensity of the PL spectra. Figure 6 clearly shows that the activation energies $E_{a2}$ of all the samples show nice agreement to $\frac{1}{2}\Delta E_i$. The remaining solid lines will be discussed later in section 3.4.4.

*3.4.3 Discussion of activation energy $E_{a1}$*. The obtained values of the small activation energy $E_{a1}$ shown in table 1 range from 20 to 22 meV for 4—6 ML InAs/InGaAlAs QDots and from 17 to 25 meV for 3—6 ML InAs/InAlAs QDashes. From this we can infer that the value of $E_{a1}$ is almost independent of the barrier materials. The mechanism behind this small activation energy of QDots and QDashes cannot be directly connected to some optical gap because it is too low to activate the carriers to the WL or barrier from the QDs ground states (GS); it rather coincides with typical p-shell excited states [41].

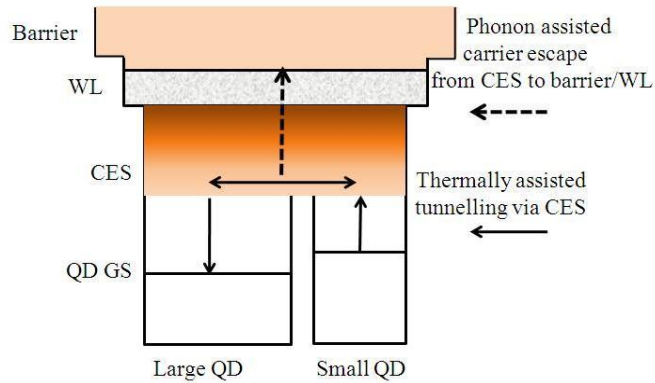

**Figure 7:** Schematic illustration of the conduction band structure exemplifying two QDs of different size and energy and the related thermally activated carrier escape and tunnelling processes from high energy to low energy QDs, as well as the common features CES, WL and barrier.

Accordingly, non-resonant phonon-assisted tunnelling of holes between QDs in adjacent stacked layers through coupled excited states (CES) in the valence-band is discussed in ref. [42]. On the other hand, also non-resonant exciton tunnelling between laterally coupled QDs in single lateral QD molecules is discussed, with the focus on the most likely transitions between indirect and direct excitons, which corresponds to an effective hopping of the electron [43]. In our present highly dense, laterally coupled QD ensembles, it is most realistic to assume that the QD excited states of neighbouring QDs are coupled, which then results in quasi-continuous CES between QD GS and WL/barrier. These CES exhibit a rather high state density, which is even increasing the closer the energy approaches the WL/barrier. All relevant states and possible charge transfer processes are illustrated in the schematic diagram, which is simplified to only the conduction band (figure 7). Despite this simplification in the illustration, a comparable electronic structure is expected for holes as well as for electron-hole pairs or excitons. Direct tunnelling between the exciton GS of neighbouring QDs is not probable due to the fairly large distances between the wave functions' centres of mass (> 50 nm). However, coupling of the excited states is much more probable because of larger extensions of the respective SQD wave functions and a reduced remaining height of the effective tunnelling barrier. Therefore, the formation of exciton CES should be possible. The thermally activated charge carrier and exciton transfer from QDs with higher GS to QDs with the lower GS via CES is the dominant tunnelling process as it leads to a minimization of the total energy. It becomes possible at relatively low temperatures, which correspond to a phonon-mediated excitation from QD GS to CES. There might be a possibility of direct radiative recombination from CES, however no relevant PL features could be observed. This observation of no CES PL together with the expected fast relaxation back to favourably lower energy QDs leads to the conclusion of a very short lifetime of theses states which is dominated by non-radiative decay. Besides the relaxation of CES to QD GS these states can also decay by non-radiative escape to the WL/barrier assisted by acoustics phonons, which is represented by the dashed arrow in figure 7. This escape process is assigned to the small intensity reduction at intermediate temperatures in all measurements and the corresponding smaller values of the activation energies, $E_{a1}$, and the corresponding coefficients $B_1$ [44]. It shall be recalled here, that all fitted $B_1$ values are comparable and small. This indicates that the aforementioned carrier escape from the CES to the WL/barrier is not very efficient and that a lot of relaxation to the QD GS occurs, which corresponds to the (re)capture processes mentioned in the definition of the $B$ coefficients (sec. 3.4.1).
A more detailed discussion of the CES mediated carrier transfer follows in the next section.

*3.4.4 Thermally assisted exciton transfer and QD repopulation.* The temperature dependent PL peak energy and PL line width are analysed in order to examine the carrier redistribution among QDs in more detail. Exemplarily 4 ML QDots and QDashes are selected for the following discussion. Figures 8 (a) and (d) illustrate the temperature dependence of the luminescence peak energy of 4 ML QDots and QDashes, respectively. The respective PL peak energies were extracted from the PL temperature series (figures 4 and 5) and are approximated for both the low and high temperature sides using Varshni empirical expression

$$E(T) = E_0 - \frac{\alpha T^2}{\beta + T} \, . \quad (3)$$

Where $E_0$ is the 0 K energy gap, $T$ is the temperature in K, $\alpha$ and $\beta$ are the fitting parameters. The two green dashed lines were calculated for guiding our eyes to the changes of energies, using the equation (3) with the parameters $\alpha$ = 0.27 meV/K and $\beta$ = 94 K [45]. This comparison clearly unveils that at intermediate temperatures between 70 K and 150 K the PL peak energies of both samples show an enhanced red shift. Similar red-shifts have also been observed for the other samples of different QD sizes that have been discussed earlier in this work (not shown here).

The origin of this anomalous red shift of the PL peak energy can be explained using the normalized PL spectra shown in figures 8 (b) and (e). As highlighted by the arrows, the spectral variations show that the PL ensembles narrow from the higher energy side. This backs up the previously suggested carrier transfer/tunnelling process from high energy (smaller sized) QDs to low energy (bigger sized) QDs via the CES.

In figure 8 (c) a decrease of PL FWHM is observed between 70 and 140 K shown for 4 ML QDots. The FWHM of the 4 ML QDot ensemble PL reduces by ~35% with temperature and reaches its minimum at 140 K. For higher temperatures the FWHM again starts to increase, which is attributed to broadening mechanisms such as electron-phonon interaction [46]. A similar reduction of the FWHM is found between 80 and 180 K for the 4 ML QDashes (figure 8 (f)). This clear trend of a FWHM reduction with increasing temperatures coincides well with the observed PL peak shift and also supports the suggestion of carrier redistribution from high to low energy QDs.

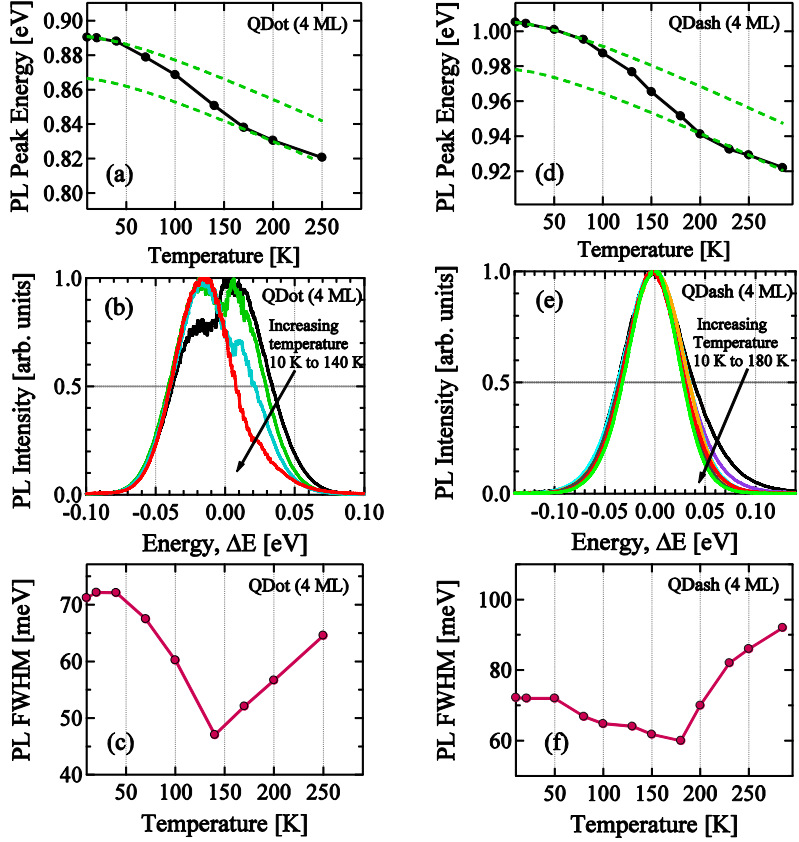

**Figure 8: (a)** Temperature dependence of the energy peak position of 4 ML QDots obtained from Gaussian fits to the spectra. The two parallel dashed lines are the calculated by using equation (3) ($\alpha$ = 0.27 meV/K and $\beta$ = 94 K). **(b)** Spectral variation of 4 ML QDots. Each spectrum is normalized to its maximum intensity and to its energy. **(c)** Temperature dependence of the 4 ML QDot PL FWHM obtained from Gaussian fits to the spectra. **(d)** Temperature dependence of the energy peak position of 4 ML QDashes, displayed in the same way as in **(a)**. **(e)** Spectral variation of 4 ML QDashes. **(f)** Temperature dependence of the 4 ML QDash PL FWHM.

This additional aforementioned PL red-shift with increasing temperatures leads to a modification of the definition of optical energy separation, $\Delta E_i$, which has been introduced in sec. 3.3 (and figure 6). To obtain a more precise estimation of $\Delta E_i$, the temperature dependent peak shifts should be included in the estimation of the QDot and QDash transition energies. For thermally activated escape processes, especially the second one, which is labelled by the index 2, mainly the higher temperature energy separations are relevant. Therefore, the values for the PL peak energies are corrected by the additional red-shift, which corresponds to the energy difference between the two dashed guide lines in figure 8. The solid lines, which highlight $\Delta E_i$ and which are displayed in figure 6, are based on this consideration. The good agreement with the measured activation energies supports the proposed carrier escape mechanism with $v_i$ = ½ for both QDots and QDashes. This mechanism suggests that the initial escape of one carrier changes the potential landscape sufficiently to dynamically induce the immediate escape of the other one. Consequently, we consider electron-hole pairs to escape the QDs on average with their concentrations remaining equal both within the QDs and the WL or barrier. We interpret this mechanism as quasi-correlated carrier escape. This somehow resembles the well-known result for the equilibrium intrinsic carrier concentration $n_i \sim \exp[-E_g/(2kT)]$, where $E_g$ is the energy gap. It shall be noted that the barrier layer is considered for QDots for the energy difference, $\Delta E_i$, and the WL is considered for the QDashes. This underlines the influence of the WL on the carrier escape processes and thus on the PL temperature dependence.

## 4. Single-QDash PL

In the final step we aimed for realizing a practical way to address single QDashes with PL energies of around 0.799 eV (1.55 μm). In order to extract single QDash emission from the high density samples, electron beam lithography and reactive ion etching were performed to fabricate tapered sub-μm sized mesa structures with approximately 200 to 300 nm height and diameters of down to 190 nm of the QDash plane [figure 9 (a)][47]. We performed micro-PL measurement of a few QDashes embedded in a nm-sized mesa structure with a diameter of 190 nm of the QDash-containing plane under non-resonant excitation and at cryogenic temperature of 4 K. We approach particularly to 6 ML QDashes following our interest towards longer emission wavelength of 1.55 μm suitable for low loss fibre optical communication. We successfully achieve single QDashes emission line close to 1.52 μm shown in figure 9 which takes us a step forward to the vision of using these dots and dashes for quantum optical applications on the single-dot/dash level in the telecommunication band.

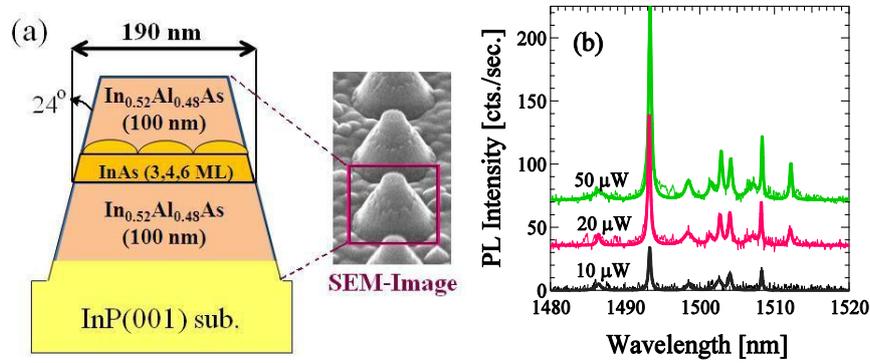

**Figure 9: (a)** Schematic illustration of a 190 nm sized mesa structure and the corresponding secondary-electron microscope (SEM) image with a 24° taper angle, a height of 230 nm and a top diameter of 100 nm which corresponds to approximately 190 nm diameter of the QDash plane. **(b)** Excitation power dependence of high resolution micro-PL spectra of single 6 ML QDashes embedded in sub-micron sized mesa structures containing emission lines from a few individual QDashes emitting near the telecommunication C band.

## 5. Summary and Conclusion

We deduce the optical gaps for InAs/InP QDot and InAs/InP QDash on different substrate orientation of (311)B and (001) from low temperature PL spectroscopy. Our experimental results confirmed the absence of a WL for the QDot samples grown on InP (311)B substrates, whereas the existence of a WL was confirmed for the QDash samples grown on InP (001) substrates. We show that the optical energy separation between QDash and WL emission is larger in comparison to the optical gap between QDot and barrier emission. Thermal activation processes of confined carriers were analyzed and the activation energies of QDashes were found larger than those of QDots for the respective number of MLs (emission energy) which reflects the achievement of higher confinement energy by the inclusion of a wider bandgap barrier material, InAlAs, in InAs QDashes samples. We confirmed a common carrier escape process for both types of samples, which holds independent of the WL presence and is attributed to a correlated electron-hole pair escape with an assigned activation energy that equals half the respective optical gap energy between QDots and barrier or QDashes and WL. A second common escape mechanism was found which corresponds to a smaller activation energy of around 20—25 meV. This mechanism is connected to a two step process of carriers being activated first to a CES, which is a unique feature for high-density QD ensembles, and second either thermalize to the barrier/WL or relax back to favorably lower energy QDs. The energy redistribution of QDs connected to this process is well reflected in an unusual shift of PL peak energy with temperature and an anomalous decrease of the ensemble FWHM. Finally, we suggest and demonstrate a way to investigate the micro-PL of nm-sized mesa structures containing small subensembles of 6 ML QDashes. We successfully demonstrated photon emission at wavelength near 1.55 μm from individual QDashes. In this way we also present a suitable way to use high density QDashes as attractive sources of single photons for application in future fibre-based quantum communication.

**Acknowledgments**

The authors thank S. Odashima for valuable comments and helpful discussion. This work was partially supported by the Grand-in-Aid for Scientific Research (A), No.21246048, HINTS by the Ministry of Education, Culture, Sports, Science and Technology, and SCOPE (Strategic Information and Communications R&D Promotion Programme) from the Ministry of Internal Affairs and Communications, Japan. CH acknowledges the Japan Society for the Promotion of Science (JSPS) for providing financial support in the form of a JSPS Fellowship for Foreign Researchers; NAJ acknowledges financial support via a MEXT scholarship.